\documentclass[%
superscriptaddress,
nofootinbib,
 amsmath,amssymb,
pra,
reprint,
]{revtex4-2}

\usepackage{graphicx}
\usepackage{dcolumn}
\usepackage{bm}
\usepackage{hyperref}

\usepackage{algorithm2e}
\usepackage{mathtools}
\usepackage{cleveref}
\usepackage{physics}
\usepackage[final,defaultcolor=red]{changes}

\newtheorem{definition}{Definition}

\makeatletter
\newcommand{\removelatexerror}{\let\@latex@error\@gobble}
\makeatother

\begin{document}

\preprint{APS/123-QED}

\title{Variational decision diagrams for quantum-inspired machine learning applications
}%

\author{Vladimir Vargas-Calderón}
\email{vvargasc@dwavesys.com}
\altaffiliation{V. V.-C. was employed by Zapata Computing Inc. for the majority of this work.}
\affiliation{D-Wave Systems, Burnaby, British Columbia, Canada}
\author{Santiago Acevedo-Mancera}%
\author{Herbert~Vinck-Posada}%
\affiliation{%
 Grupo de Superconductividad y Nanotecnología, Departamento de Física, Universidad Nacional de Colombia, Bogotá, Colombia
}%

\date{\today}

\begin{abstract}
Decision diagrams (DDs) have emerged as an efficient tool for simulating quantum circuits due to their capacity to exploit data redundancies in quantum states and quantum operations, enabling the efficient computation of probability amplitudes.
However, their application in quantum machine learning (QML) has remained unexplored.
This paper introduces variational decision diagrams (VDDs), a novel graph structure that combines the structural benefits of DDs with the adaptability of variational methods for efficiently representing quantum states.
We investigate the trainability of VDDs by applying them to the ground state estimation problem for transverse-field Ising and Heisenberg Hamiltonians.
Analysis of gradient variance suggests that training VDDs is possible, as no signs of vanishing gradients--also known as barren plateaus--are observed.
This work provides new insights into the use of decision diagrams in QML as an alternative to design and train variational ansätze.
\end{abstract}

\maketitle


\section{Introduction}

Quantum circuit simulation on classical computers is essential because it allows researchers to develop, test, and verify quantum algorithms for quantum chemistry~\cite{mazzola2024quantum}, condensed matter~\cite{Bassman_Oftelie_2021}, high-energy physics~\cite{meglio2024quantum} and many other applied fields of science and technology.
However, simulating quantum circuits is generally exponentially hard due to the rapidly growing size of quantum states as the number of qubits increases.
Despite this challenge, certain quantum circuits can be efficiently simulated on classical computers, particularly when specific structures or patterns reduce complexity~\cite{rudolph2025paulipropagationcomputationalframework}.
For instance, recent research has shown that quantum circuit simulability is deeply connected with the Lie-algebraic structure underlying the accessible quantum states of a given ansatz~\cite{goh2023liealgebraicclassicalsimulationsvariational,ragone2024lie,bermejo2024quantumconvolutionalneuralnetworks,kazi2024analyzingquantumapproximateoptimization}.

Decision diagrams (DDs) are data structures that offer a powerful approach for representing multivariable functions with important applications such as optimisation when the function is a cost function, or conditioning when the function is a probability distribution~\cite{Fargier_Marquis_Niveau_Schmidt_2014,sistla2024cflobdds,Kisa2014PSDD,Wilson2005Semiring,thierry2009hierarchical,Sanner2005AADD,Fujita1997MTBDD,Lai1996EVBDD,Tafertshofer1997FEVBDD,vinkhuijzen2024knowledgecompilationmapquantum}.
DDs can also be used to represent quantum states (which are related to probability distributions through the Born rule) and quantum gates (which are complex mappings) in a concise manner~\cite{Burgholzer2025MQTCore}.
So far, different ways of using these data structures have been proposed to simulate quantum circuits, where the decision diagram that represents the quantum state of a quantum register is updated after the action of the quantum gates found in the quantum circuit~\cite{miller2006qmdd,sistla2024weighted,hong2025limtddcompactdecisiondiagram,brand2025qsylvanparalleldecisiondiagram}.
In this direction, DDs have proven effective in verifying a wide range of quantum algorithms \cite{zulehner2018advancedsim,hillmich2020fastweaksim,hillmich2021asaccurateasneeded,burgholzer2021hybridshrodingerfeynman,burgholzer2022exploitingarbitrarypaths,grurl2023noiseaware}, even for circuits involving a large number of qubits, which demonstrates its utility in handling complex quantum computations.

\begin{figure}
    \centering
    \includegraphics[width=\linewidth]{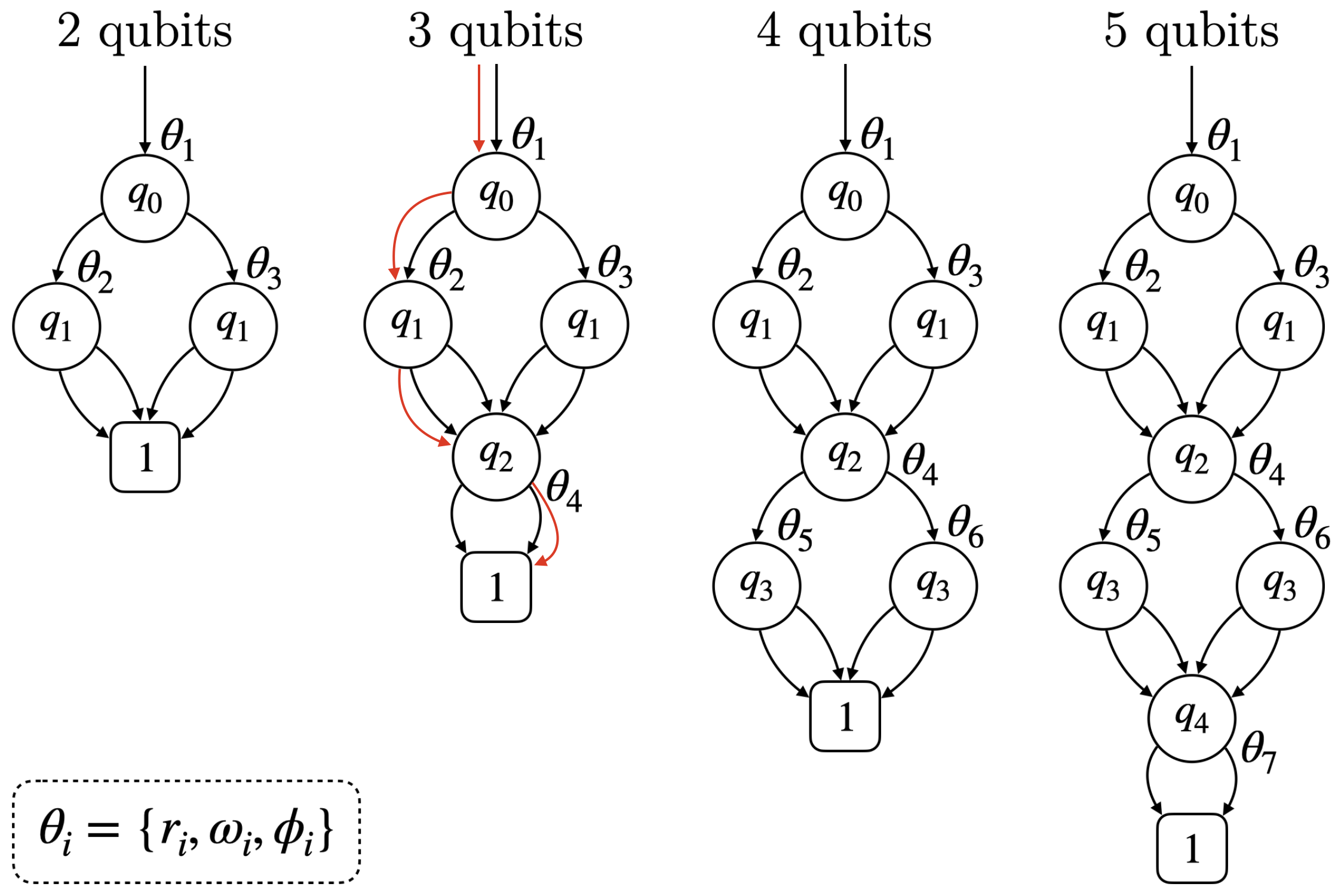}
    \caption{Schematic layout of the VDD accordion ansatz for 2-5 qubits. At the top of each diagram is the root node, and at the bottom is the terminal node. The VDD has variational parameters at each edge of the diagram. To obtain the amplitude probability of a given element of the canonical basis of the Hilbert space of a system of $n$ qubits (a bit string of length $n$), all that is needed is to take the path from the root node to the terminal node, where if a zero(one) is encountered in the bit string, the left(right) edge is taken; the probability amplitude is the product of all probability amplitudes lying on the edges of the specified path (see~\cref{def:qdd,def:vdd} for details on how the probability amplitudes are defined and computed).
    The red path is highlighted in the 3 qubit case is related to an example that is explained in the main text (cf.~\cref{eq:prob_amplitude_example}).}
    \label{fig:accordion}
\end{figure}

On the other hand, variational quantum circuits have emerged as a promising tool for using quantum computers to solve relevant problems in fields such as physics and optimisation~\cite{cerezo2021variational}.
Succinctly, variational quantum circuits are built using parameterised quantum gates that act on small subsets of qubits; these parameters are then optimised to minimise a specified loss function, typically by computing the gradients of this loss with respect to the variational parameters and then using these gradients to update the parameters themselves.
Readers unfamiliar with this field are referred to Refs.~\cite{cerezo2021variational,qi2024variational}.
Even though variational quantum circuits were initially thought to hold great promise in the practical utility of quantum computers, the reality is that, for most applications, initialisation of the variational parameters tend to yield exponentially small gradients in the number of qubits, a phenomenon coined as the presence of barren plateaus~\cite{mcclean2018barren}, limiting the scale (in terms of number of qubits) at which variational quantum circuits can be used.

Despite their relative success in quantum circuit simulation, to the best of our knowledge, there has been no research on using DDs as variational ansätze for quantum machine learning (QML) applications\footnote{The closest, yet not variational, is Ref.~\cite{tanaka2024quantumstatepreparationfree}}.
The exploration of DDs in QML is intriguing for several reasons.
While the primary goal of QML is to discover algorithms that leverage quantum advantages (a task that has so far eluded the QML community~\cite{gilfuster2024relationtrainabilitydequantizationvariational,bermejo2024quantumconvolutionalneuralnetworks}), quantum-inspired methods like DDs could offer valuable insights and alternatives to tasks such as quantum state verification.
A natural way of using DDs as variational ansätze is to use quantum circuit simulators based on DDs to simulate variational quantum circuits.
However, this approach will only be useful if the designed quantum circuit ansatz is devoid of exponentially-small gradients.
This is a trivial application of using DDs for optimisation tasks that admit variational ansätze.
Instead, if DDs are used as a quantum-inspired tool for solving quantum problems in classical computers, they might enable the development of new ansätze for problems in quantum chemistry or condensed matter physics, where other classical simulation techniques are often limited.

In this work, we introduce Variational Decision Diagrams (VDDs), a novel paradigm that combines the compactness of DDs with the adaptability of a variational ansatz.
Unlike other state-of-the-art approaches such as neural-network quantum states (NQSs)~\cite{carleo2017solving}, tensor networks such as matrix product states (MPS)~\cite{verstraete2008matrix,orus2014tensornetworks,schollwock2011mps}, and other conventional state-vector methods, VDDs rely on a complete and implicit normalised representation of quantum states through nodes and parameterised edges.
Specifically, tensor networks decompose a quantum state into connected tensors highly efficient in certain systems with low entanglement or dimensionality, whereas VDDs do not rely on explicit tensor representations.
Instead, they capture amplitude correlations through its paths.
NQSs \cite{vivas2022nqs} treat amplitudes as trainable functions in a neural network, which can capture correlations but suffer from optimisation challenges characteristic of the neural network architectures used for these variational ansätze.
For non-autoregressive neural network architectures, NQSs remain unnormalised; on the other hand, VDDs enforce normalisation constraints at each node in a straightforward manner.
Finally, representing the full state vector becomes impractical even for moderately large systems due to exponential scaling in memory requirements.
By contrast, VDDs can exploit shared substructures to maintain computational feasibility, thus combining the strengths of tensor networks--the ability to sparsely connect subsystems of a quantum system--and autoregressive NQSs--the ability of exploiting autodifferentiation~\cite{sharir2020deepautoregressive}.

In this work, we make three main contributions: (1) we propose VDDs for the first time as a quantum-inspired parameterised structure to represent quantum states; (2) we demonstrate that, using a particular setup of VDDs that we call the Accordion ansatz, VDDs do not exhibit the barren plateau phenomenon, showing a non-exponential scaling of gradient variance with the number of qubits; and (3) we validate their effectiveness by applying them to ground state estimation for various Hamiltonians.

The remainder of this paper is organised as follows. In~\cref{sec:methods}, we formally introduce the concept of VDDs, defining their particular parameterised structure.
We then introduce the ground state estimation problem in~\cref{sec:gse}.
\Cref{sec:trainability} focuses on the methods used for optimisation and the computation of gradients, showing the absence of barren plateaus.
In~\cref{sec:results}, we present the performed experiments that numerically demonstrate that VDDs using the Accordion ansatz can successfully approximate ground states of relevant physical models.
We discuss strenghts, weaknesses and future work in~\cref{sec:discussion}.
Finally, we provide conclusions of our work in~\cref{sec:conclusions}.

\section{Methods}\label{sec:methods}

A DD that represents a quantum state is a binary directed acyclic multigraph (BDAMG) that has a root node with no parent, and a terminal node with no children.
The root node has only one outward edge, but all other nodes (except for the terminal node) have two outward edges, since it is a binary multigraph.
In~\cref{fig:accordion}, examples of VDDs are depicted, where the parent node is not shown, but is situated on top of every diagram, and the terminal node is represented by the square at the bottom of every diagram.
Nodes (except the root and the terminal nodes) represent qubit indices, and edges hold information about probability amplitudes.
Then, the two edges spawning from a node are distinguishable not only from their probability amplitude information, but because they also hold a pointer to an element of the qubit basis, i.e., a pointer to $\ket{0}$ or $\ket{1}$.
In~\cref{fig:accordion}, the left(right) outward edges of each node point to $\ket{0}$($\ket{1}$).

The two outward edges of every node can point towards a single child or towards two different children, where a precedence relationship must be maintained between the connected nodes, meaning that given a node corresponding to the $q_i$ qubit can only be connected to a node corresponding to the $q_{i+1}$ qubit, and for this reason, the graph cannot be cyclic.
Thus, indices of the nodes when traversed in any path from the root node to the terminal node are written in increasing consecutive order.
This implies that any path from the root to the terminal node will always traverse $n$ nodes for a DD that represents the state of a system of $n$ qubits.
Similarly, each element of the basis of the $n$-qubit system can be associated with a traversed path.
For example, in the case of 3 qubits, a path in red is shown.
This path corresponds to the basis state $\ket{0,0,1}$; reading from left to right, when the first node is encountered, the first zero in the basis state indicates that the path takes the left edge, when the second node is encountered, the second zero indicates that the path goes to the left edge of that node, and when the final node is encountered, the third element in the basis state, i.e., the 1, indicates that the path takes the right edge.
The DD structure allows us to compute the amplitude probability of that basis state, i.e., the quantity $\psi(000) = \braket{\psi}{0,0,1}$.
How this is done will be formalised in~\cref{def:qdd}.
We show the probability amplitude for this particular example later on in~\eqref{eq:prob_amplitude_example}.
In general, a path from the root node to the terminal node will be associated with the basis state $\ket{b_1,b_2,\ldots,b_n}$, where $b_i\in\{0, 1\}$.

With these notions, we can define a quantum state of $n$ qubits with a BDAMG as follows
\begin{definition}
    \label{def:qdd}
    A quantum decision diagram (DD) defines a wavefunction $\psi:\mathbb{B}^n\to\mathbb{C}$ that represents, in the computational basis, the quantum state of a system of $n$ qubits.
    Since every node in the DD has two outward edges (except from the root and terminal nodes) an edge can be identified by a pointer $b\in\{0,1\}$, and we denote the edge as edge$(\text{node}, b)$ and its corresponding amplitude probability as edge$(\textrm{node}, b)_p$.
    The amplitude probability of the only outward edge from the root node is edge$(\textrm{root node})_p$, and corresponds to the global phase of the quantum state.
    Further, for any edge $e$, we can uniquely identify the node to which the edge is pointing, and we denote it by $\text{child}(e)$.
    
    The following protocol defines how the value of $\psi$ is accessed at a particular bit string $\vb*{b}=(b_1,\ldots,b_n)$:
    
    \begin{myalgorithm}
    \removelatexerror
        \SetAlgoLined
        \KwData{A bit string $\vb*{b}$.}
        \KwResult{The probability amplitude $\psi(\vb*{b}) = (\bra{b_1}\otimes\cdots\otimes\bra{b_n})\ket{\psi}$.}
        \Begin{
            node $\leftarrow$ root node\\
            $\varphi \leftarrow$ edge$(\textrm{node})_\text{p}$\\
            \For{$i\leftarrow n$ \KwTo $1$}{
                $e \leftarrow \text{edge}(\text{node}, b_i)$\\
                $\varphi \leftarrow \varphi \times e_\text{p}$\\
                node $\leftarrow$ child$(e)$
            }
            $\psi(\vb*{b}) \leftarrow \varphi$
        }
    \end{myalgorithm}
    
    The wavefunction will be normalised to 1 if $|$edge(root node)$_p|=1$ and $|$edge(node, 0)$_p|^2 + |$edge(node, 1)$_p|^2=1$ for all the other nodes, except for the terminal node.
\end{definition}

In variational quantum circuits, the wavefunction $\psi$ is parameterised through parameterised quantum gates.
In analogy, we define a variational decision diagram as follows:

\begin{definition}\label{def:vdd}
    A VDD is a DD where the probability amplitudes of each pair of outward edges spawning from nodes that are not the root or terminal nodes are parameterised by three real parameters $(r, \omega, \phi)$ such that
    \begin{align}
        \text{edge}(\text{node}, 0)_\text{p} = re^{i\omega}\;\text{and}\;\text{edge}(\text{node}, 1)_\text{p} = \sqrt{1-r^2}e^{i\phi},\label{eq:nodes_weights}
    \end{align}
    where $r\in[0, 1]$.
\end{definition}

Notice that \cref{def:vdd} ensures that the quantum state represented by the VDD is always normalised to 1.

Continuing with the example of the red path in~\cref{fig:accordion}, the probability amplitude $\psi(001) = \braket{\psi}{0,0,1}$ is given by the product of the probability amplitudes lying on each segment of the path, i.e.,
\begin{align}
    \underbrace{r_1e^{i\omega_1}}_{\text{First segment}}\times\underbrace{r_2e^{i\omega_2}}_{\text{Second segment}}\times\underbrace{\sqrt{1-r_4^2}e^{i\phi_4}}_\text{Third segment},\label{eq:prob_amplitude_example}
\end{align}
where we have obviated the global phase associated to the outward edge of the root node.

\section{Ground state estimation}\label{sec:gse}

Given a Hamiltonian $H$, we minimise its expected value with respect to a quantum state described by a VDD\footnote{We refer to the expectation value also as ``the loss function''.}, i.e., we solve
\begin{align}
    \min_{\vb*{\theta}} \expval{H}{\psi_{\vb*{\theta}}},\label{eq:minproblem}
\end{align}
where $\psi_{\vb*{\theta}}$ is the wavefunction described in~\cref{def:qdd} for a DD realised via a VDD, as described in~\cref{def:vdd}.
In other words, $\vb*{\theta}$ refers to the collection of all variational parameters of the VDD described in~\cref{def:vdd}.
The minimisation in~\cref{eq:minproblem} is carried out with stochastic gradient descent, for which the gradient
\begin{align}
    \grad_{\vb*{\theta}} \expval{H}{\psi_{\vb*{\theta}}} \label{eq:gradient_of_hamiltonian}
\end{align}
needs to be computed.
In order to estimate the components of the gradient in~\cref{eq:gradient_of_hamiltonian}, one straightforward approach involves constructing a state vector representation of the VDD.
The gradients, as depicted in~\cref{eq:gradient_of_hamiltonian}, can be computed by recognising that $\braket{\vb*{b}}{\psi_{\vb*{\theta}}}$ is a differentiable function of $\vb*{\theta}$. In fact,
\begin{align}
\begin{aligned}
    \braket{\vb*{b}}{\psi_{\vb*{\theta}}} ={}& \text{edge}(\text{node}_{p_1}, b_n)_\text{p} \text{edge}(\text{node}_{p_2}, b_{n-1})_\text{p}\\
    &\text{edge}(\text{node}_{p_3}, b_{n-2})_\text{p}\cdots\text{edge}(\text{node}_{p_n}, b_1)_\text{p},
\end{aligned}\label{eq:element_of_vdd}
\end{align}
where this differentiable function has been directly obtained from~\cref{def:qdd}. The numbers $p_i$ in~\cref{eq:element_of_vdd} index the nodes in the DD that are traversed for the particular bit string $\vb*{b}$ under consideration.
In other words, one can look at the protocol in~\cref{def:qdd} as a protocol that accumulates the product of probability amplitudes of the form $\alpha_i^{\beta_i}$ (cf.~\cref{eq:nodes_weights}) along the path that is determined by the bit string under consideration, where $i$ precisely indexes the nodes along this path.
The specific form of $\alpha_i^{\beta_i}$ is determined by the parameters $(r_i,\omega_i,\phi_i)$ defined in~\cref{def:vdd} for a particular node $i$.

As a result, it is straightforward to automatically differentiate the expected value of the Hamiltonian with respect to the parameters of a variational state vector whose components are $\braket{\vb*{b}}{\psi_{\vb*{\theta}}}$.
This can be done using widely adopted machine learning libraries such as PyTorch~\cite{paszke2019pytorchimperativestylehighperformance} or JAX~\cite{jax2018github}.
However, this method requires the explicit calculation of the state vector, which becomes inefficient (due to exponential scaling) as the number of qubits increases.
For the purposes of this paper, where the required calculations to show our findings involve a small number of qubits (on the order of 10 qubits), we rely on building the whole state vector.
We anticipate that our results will be applicable to larger numbers of qubits as well.
To this end, in \cref{sec:vmc}, we introduce a scalable approach to compute the required gradients for large systems (\cref{eq:gradient_of_hamiltonian}) using variational Monte Carlo.

We consider different Hamiltonians such as
\begin{align}
    H_1=Z_1Z_2,\label{eq:z1z2}
\end{align}
where $Z_i$ is the Pauli $Z$ matrix applied to the $i$-th spin.
This model is considered, despite its simplicity, as it is prototypical to show the barren plateau phenomenon in the seminal barren plateau paper (Ref.~\cite{mcclean2018barren}).
We also consider the transverse-field Ising model (TFIM) defined on $N$ spins
\begin{align}
    H_2 = \sum_{\langle i, j \rangle}Z_iZ_j + g \sum_{i=1}^NX_i,\label{eq:ising}
\end{align}
where $\langle i,j\rangle$ defines all neighbouring pairs in a 1D chain of spins, $g$ is the relative strength of the transverse field with respect to the spin-spin coupling and $X_i$ is the Pauli $X$ matrix applied on the $i$-th spin.
Finally, we consider the XYZ Heisenberg Hamiltonian
\begin{align}
    H_3 &= \sum_{\langle i, j \rangle} (X_iX_j + Y_iY_j + Z_iZ_j)\label{eq:heisenberg}
\end{align}
where $Y_i$ is the Pauli $Y$ matrix applied on the $i$-th spin.

Problems different from ground state estimation can also be tackled, as the only requirement for gradient-descent optimisation to be carried out is to be able to estimate gradients of loss functions with respect to the VDD's parameters.

\begin{figure*}[t]
    \centering
    \includegraphics[width=0.85\textwidth]{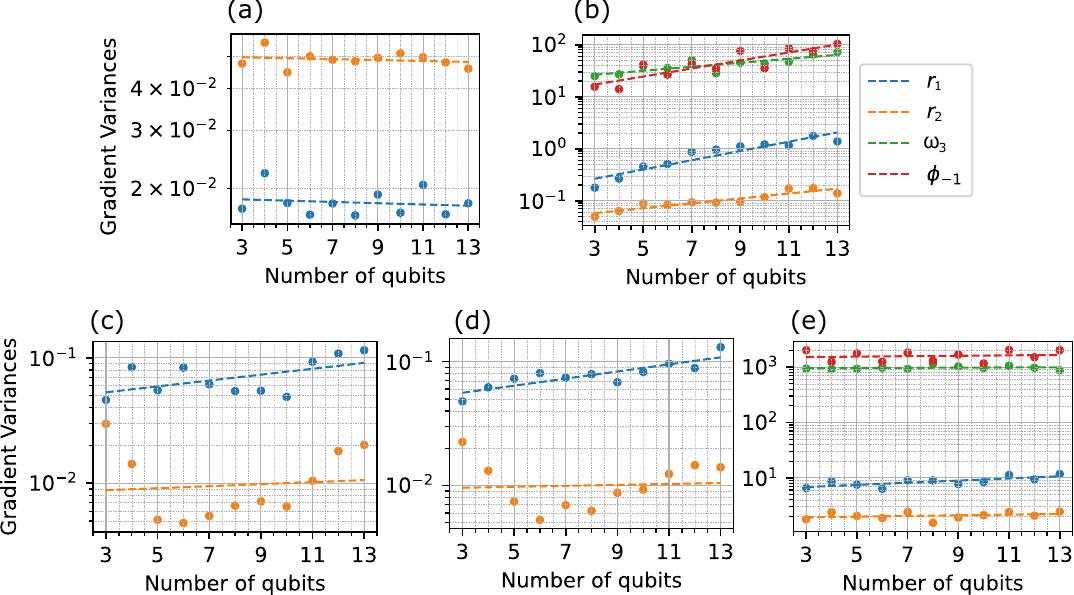}
    \caption{Gradient variances, computed with~\cref{eq:variance_of_gradients}, averaged over random values of some parameters of the accordion ansatz for the expected value of the following Hamiltonians: a) $Z_1Z_2$; b) Heisenberg with $J_x=J_y=J_z=1.0$; TFIM with $g=0.0$ (ordered phase), d) $g=1.0$ (gapless phase) and e) $g=10.0$ (disordered phase).
    Linear fits are also shown. $\phi_{-1}$ refers to the $\phi$ parameter of the last edge of the VDD.}
    \label{fig:variances}
\end{figure*}

\section{Trainability}\label{sec:trainability}

In the context of quantum circuits, i.e., when $\psi_{\vb*{\theta}}$ is realised by a variational quantum circuit, it has been shown that the minimisation problem in~\cref{eq:minproblem} can be solved in practice only when the variance of a component of the aforementioned gradient decays sub-exponentially with respect to the number of qubits in the circuit ~\cite{mcclean2018barren}.
The reason for this is that gradient descent algorithms update the parameters with a step size usually proportional to the gradient~\cite{wang1989training}, meaning that if the gradient does not vanish exponentially fast in the number of qubits, the number of steps to convergence will also not be exponential in the number of qubits.
This property is known as trainability, and it shows the ability that gradient descent-based algorithms have to navigate the space of parameters.
Lack of trainability is usually metaphorically described as the loss function landscape being flat like a barren plateau, and thus, it is said that the loss function landscape has barren plateaus.
It is important to highlight that trainability does not guarantee convergence towards a global minimum ~\cite{anschuetz2022traps}, as the loss function landscape might contain several local minima that can trap the gradient descent algorithm.

Similarly, we use the scaling of the $j$-th component of the variance of the gradient, i.e.,
\begin{align}
    \text{Var}\left(\pdv{\theta_j}\expval{H}{\psi_{\vb*{\theta}}}\right)\label{eq:variance_of_gradients}
\end{align}
as a function of the number of qubits to assess trainability of a given ansatz.
In our case, $\psi_{\vb*{\theta}}$ is a VDD.
To avoid confusion, in~\cref{eq:variance_of_gradients}, $j$ indexes a single parameter of the complete collection of parameters $\vb*{\theta}$.
It does not refer to the collection of three parameters corresponding to the $j$-th node of a VDD (cf.~\cref{fig:accordion}).

We use the Hamiltonians in \cref{eq:z1z2,eq:ising,eq:heisenberg} to assess the scaling of~\cref{eq:variance_of_gradients}.
We assess the scaling of \cref{eq:variance_of_gradients} using a VDD defined by an ansatz we term the accordion ansatz.
We highlight the fact that results in this section are valid for this ansatz, and other DD structures may have different trainability properties.

\subsection{Accordion ansatz}
\label{sec:accordion}
We introduce the VDD accordion ansatz as a VDD composed by alternating one- and two-node levels (except for the terminal node).
This means that always, the first qubit will be represented by the first level, which consists of one node, the second qubit will be represented by the second level, which consists of two nodes, and so on.
\Cref{fig:accordion} shows a schematic representation of the graph structure with its corresponding parameters, where $\theta_i$ is the set of parameters $\{r_i,\omega_i,\phi_i\}$ for each node of the VDD, as is mentioned in \cref{def:vdd}.
\added{Thus, the number of trainable parameters of the Accordion ansatz is $3\lfloor3n/2\rfloor$.}

\added{The accordion ansatz is a natural departure from a product state ansatz, which can be diagrammed as a VDD simply as having one node per level.
Therefore, the accordion ansatz naturally parameterises a dimer product state ansatz of the form $\ket{\psi_{1,2}}\otimes\ket{\psi_{3,4}}\otimes \cdots$, where each $\ket{\psi_{I}}$ is an arbitrary quantum state of, at most, two-qubits indexed by a pair $I$ (in the case of even $n$, all kets are arbitrary states of two qubits, but in the case of odd $n$, the last ket is a single-qubit state).}

In order to assess the trainability of the VDD accordion ansatz, we measure the variance of different components of the gradient of the loss function with respect to the variational parameters.
This variance is shown in \cref{fig:variances}, from where we can see that it does not decay exponentially fast with respect to the number of qubits.
Thus, the accordion ansatz does not show the barren plateau phenomenon for the Hamiltonians under consideration, asserting its trainability.
This behaviour holds for parameters that refer to the relative amplitude between the $\ket{0}$ and $\ket{1}$ states for each qubit ($r$ parameters) as well as the parameters that fix the relative phases between the states ($\omega$ and $\phi$ parameters).
\added{The accordion ansatz is expected to be free of barren plateaus due to its limited expressivity, as it effectively corresponds to a dimerised product-state ansatz.
By contrast, more general VDD constructions---up to and including the fully universal case where each node splits into two nodes in the next level---are capable of representing states with longer-range entanglement.
Intuitively, as the expressivity of the VDD increases, at the expense of a larger number of trainable parameters, the influence of any individual parameter on the energy gradient is expected to lower.
This mirrors the behaviour observed in deep parametrised quantum circuits~\citep{ragone2024lie}, where enhanced expressivity is often accompanied by unfavourable scaling of gradient variances.}

\section{Results}\label{sec:results}
In this section, we present results on the quality of ground state estimation on the problems defined in \cref{eq:z1z2,eq:ising,eq:heisenberg}. Since our approach to optimisation relies on concretely calculating the state vector represented by the VDD, we can automatically differentiate any function involving this state vector. In particular, we minimise the difference between the expected value of the Hamiltonian with respect to the VDD and the actual ground energy of the given system. 

\begin{figure*}[t]
    \centering
    \includegraphics[width=1\textwidth]{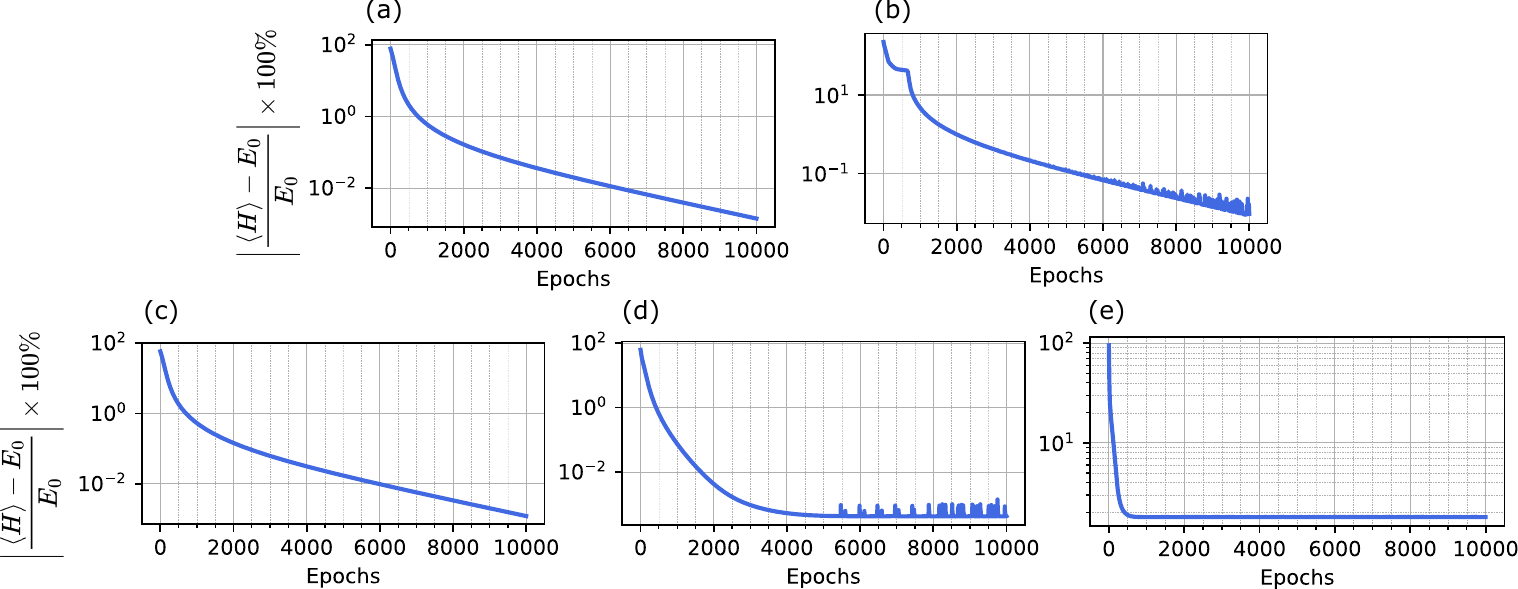}
    \caption{Percentual energy error curves as a function of the epoch number for a system with 10 qubits for the following Hamiltonians: a) $Z_1Z_2$, b) Heisenberg with $J_x=J_y=J_z=1.0$, c) TFIM with $g=0.0$ (ordered phase), d) TFIM with $g=1.0$ (gapless phase) and e) TFIM with $g=10.0$ (disordered phase). The optimised loss function is the difference between the expected value of the corresponding Hamiltonian and its ground state energy. $E_0$ is the true ground energy of each Hamiltonian.}
    \label{fig:lossfn}
\end{figure*}

As previously mentioned, we calculated the variance of the gradient of the loss function with respect to each parameter of the VDD varying the number of qubits of the graph structure defined by the accordion ansatz, using 100 random parameter vectors with different random seeds.
The corresponding results obtained in these experiments are shown in~\cref{fig:variances}, where only a subset of the parameters are shown, as they qualitatively represent the behaviour of the parameters that are not shown.
In some cases, some parameters exhibit near zero variance, where the gradients are not sensitive to some parameter changes for some of the Hamiltonians under consideration.
Such is the case of $H_1$, in~\cref{fig:variances}(a), where it is clear that the parameters $\omega_3$ and $\phi{-1}$, useful to define the phase structure of the VDD state, do not intervene in the expected energy gradient.

As per the approximation of ground state energies, \cref{fig:lossfn} shows the optimisation curve for each Hamiltonian considered in a system of 10 qubits using the difference between the estimated and the actual ground energy as the loss function to minimise.
For these experiments, we have used the Adam \cite{kingma2017adam} optimiser with a learning rate of 0.01 and 10000 epochs.
In all cases, the VDD's variational parameters are initialised randomly.

All experiments shown in~\cref{fig:lossfn}, but panel (e), show progressive minimisation of the loss function.
For this panel, the ground state of the TFIM model for $g=10$ is close to $\ket{+,\ldots,+}$, which is the ground state the VDD converges to, where the $r$ parameter of every node is $\frac{1}{\sqrt{2}}$, and all other parameters are zero.
The larger the transverse field is, the closer the true ground state is to $\ket{+,\ldots,+}$, and the better the VDD is able to approximate such state.
Strangely, the VDD is not able to approximate the true ground state energy of the TFIM ($E_0$), for moderate strengths of the transverse field in the disordered phase.
For the TFIM, we have observed that for $g>10$ the error roughly satisfies $\abs{(\expval{H}-E_0)/E_0}\propto{1/g}$.
\added{This is expected because, in the disordered phase, the $ZZ$ term in the TFIM induces short-range correlations that decay in strength with larger $g$.
Such a quantum state is not representable by an arbitrary dimer product state (see~\cref{sec:accordion}), which is why the VDD has difficulty representing it, but improves with increasing transverse field strength $g$.}

\section{Discussion}\label{sec:discussion}
The numerical results presented in~\cref{sec:results} showcase the viability of using VDDs for ground state estimation problems.
We argue that VDDs can offer a more compact or efficient representation than tensor networks, NQSs and other full state-vector methods, especially when the structure of the problem can be reduced using this representation, or when explicit wavefunction normalisation is needed.
Nonetheless, this comes at a significant cost: the design of the ansatz must match the correlations of the system.
This means that fixing an ansatz already proves VDDs inflexible when trying to exploit correlations between qubits far apart from each other.
However, this need not mean that it is impossible to propose VDD architectures a priori.
For example, tensor network architectures have been designed inspired in the arrangement of qubits on a chip to simulate results from that chip~\cite{tindall2024efficient,patra2024efficient}.
Other geometries could exhibit advantages in capturing long-range entanglement or highly correlated configurations.
However, they could also introduce additional parameters that complicate the optimization landscape.
Thus, a possible direction for future research is to explore alternative VDD ansätze for solving more complex problems\added{, which involve studying the expressivity of these VDD ansätze in the light of approximating Haar-random states with the $t$-design approximation theory~\citep{nakata2021quantum,belkin2024approximate}.}

Finally, although this work only treats the ground state estimation problem, VDDs can be used to solve other quantum machine learning tasks, such as classification\added{,  regression or generation}, where the loss function can be different, e.g. portfolio optimisation~\cite{alcazar2024enhancing}.
\added{More precisely, VDDs can be used in machine learning tasks that rely on loss functions that require the explicit evaluation of the probability of a bit string under the model.
For example, binary classification can be achieved by training a VDD with a training dataset $\{(\vb*{b}_i, \ell_i)\}_{i=1}^N$, where a bit string $\vb*{b}_i$ has label $\ell_i\in\{0, 1\}$, by minimising the binary cross entropy, given by
\begin{align}
\begin{aligned}
    \text{BCE}(\vb*{\theta}) = -\frac{1}{N}\sum_{i=1}^N&\left(\ell_i \log \abs{\braket{\vb*{b}_i}{\psi_{\vb*{\theta}}}}^2\right. \\
    &+ \left.(1-\ell_i) \log (1-\abs{\braket{\vb*{b}_i}{\psi_{\vb*{\theta}}}}^2)\right).
\end{aligned}
\end{align}
Similarly, a VDD can be used for generative modelling purposes by training it to minimise the Kullback-Liebler divergence between the empirical uniform distribution given by the training data and the model distribution, i.e.,
\begin{align}
    \text{KL}(\vb*{\theta}) = - \frac{1}{N}\sum_{i=1}^N\log \abs{\braket{\vb*{b}}{\psi_{\vb*{\theta}}}}^2.
\end{align}
Generating new bit strings can be done autoregressively as specified in~\cref{sec:autoregressive_sampling}.}

Future efforts involve investigating the efficacy of VDDs to solve these other tasks.

\section{Conclusions}\label{sec:conclusions}

We have introduced Variational Decision Diagrams (VDDs), a novel quantum-inspired structure that combines the compactness and efficiency of decision diagrams with the flexibility of variational methods for solving problems in many domains.
Our key contributions include (i) proposing the VDD framework for representing and optimising quantum states with respect to some loss function, (ii) demonstrating the absence of barren plateaus in the Accordion ansatz for relevant Hamiltonians, and (iii) validating the efficacy of VDDs in ground state energy estimation for various Hamiltonians.
These results highlight the potential of VDDs as a powerful tool for simulating quantum systems on classical computers, particularly in scenarios where traditional methods face scalability challenges.

Our findings show that VDDs can efficiently represent non-trivial quantum states, offering a promising alternative to existing methods such as tensor networks and neural-network quantum states (NQSs).
Unlike tensor networks, which rely on explicit tensor decompositions, VDDs capture amplitude correlations through their paths, enabling a more compact representation of quantum states.
Additionally, VDDs enforce normalisation constraints in a straightforward manner, addressing a key limitation of many non-autoregressive NQSs, which often struggle with normalisation and optimisation challenges.
This makes VDDs particularly suitable for problems in quantum chemistry, condensed matter physics, and other domains where accurate and efficient state representation is critical.

However, our work also reveals certain limitations.
While the Accordion ansatz demonstrates effectiveness in avoiding barren plateaus and approximating ground states, it struggles to capture subtle state structures in certain regimes, such as moderate transverse fields in the disordered phase of the TFIM.
This suggests that the design of the ansatz plays a crucial role in the performance of VDDs, and more complex arrangements may be necessary to fully exploit the correlations present in physical systems.
Future research could explore parameter sharing to leverage symmetries in physical systems, as well as alternative VDD architectures that better capture long-range entanglement and highly correlated configurations.

Beyond ground state estimation, VDDs hold promise for a wide range of quantum machine learning (QML) tasks, including classification, regression, and optimisation problems.
For instance, VDDs could be applied to portfolio optimization, quantum state verification, and other tasks where classical simulation of quantum systems is required.
The ability to combine the strengths of decision diagrams with variational methods opens new avenues for developing quantum-inspired algorithms that can tackle problems currently beyond the reach of classical techniques.

In conclusion, VDDs represent a significant step forward in the quest for efficient classical simulation of quantum systems.
By bridging the gap between decision diagrams and variational methods, VDDs offer a versatile and scalable framework for representing and optimising quantum states.
While challenges remain, particularly in designing ansätze that can capture complex correlations, the potential applications of VDDs in quantum chemistry, condensed matter physics, and QML are vast, just as well-established methods such as tensor networks.
Overall, VDDs provide a robust and flexible platform for variational quantum simulation.

\section*{Acknowledgements}
S.~A.-M. and V.~V.-C. are grateful to the project ``Aprendizaje de Maquina para Sistemas Cuánticos'' with HERMES code 57792. H.~V.-P. thanks the project ``Ampliación del uso de la mecánica cuántica desde el punto de vista experimental y su relación con la teoría generando desarrollos en tecnologías cuánticas útiles para metrología y computación cuántica a nivel Nacional'' with BPIN code 2022000100133 from SGR of MINCIENCIAS, Gobierno de Colombia.
All authors acknowledge Juan~E.~Ardila-García and Nicolas Parra-A. for their valuable feedback on our work.

\section*{Data Availability}
Data and code is available upon reasonable request to the authors.

\bibliography{refs,refs2}

@misc{zulehner2018advancedsim,
    title={Advanced Simulation of Quantum Computations}, 
    author={Alwin Zulehner and Robert Wille},
    year={2018},
    eprint={1707.00865},
    archivePrefix={arXiv},
    primaryClass={quant-ph},
    url={https://arxiv.org/abs/1707.00865},
}

@inproceedings{hillmich2020fastweaksim,
    title={Just Like the Real Thing: Fast Weak Simulation of Quantum Computation},
    url={http://dx.doi.org/10.1109/DAC18072.2020.9218555},
    DOI={10.1109/dac18072.2020.9218555},
    booktitle={2020 57th ACM/IEEE Design Automation Conference (DAC)},
    publisher={IEEE},
    author={Hillmich, Stefan and Markov, Igor L. and Wille, Robert},
    year={2020},
    month=jul
}

@inproceedings{hillmich2021asaccurateasneeded,
    author={Hillmich, Stefan and Kueng, Richard and Markov, Igor L. and Wille, Robert},
    booktitle={2021 Design, Automation \& Test in Europe Conference \& Exhibition}, 
    title={As Accurate as Needed, as Efficient as Possible: Approximations in DD-based Quantum Circuit Simulation}, 
    year={2021},
    volume={},
    number={},
    pages={188-193},
    doi={10.23919/DATE51398.2021.9474034}
}

@inproceedings{burgholzer2021hybridshrodingerfeynman,
    title={Hybrid Schrödinger-Feynman Simulation of Quantum Circuits With Decision Diagrams},
    url={http://dx.doi.org/10.1109/QCE52317.2021.00037},
    DOI={10.1109/qce52317.2021.00037},
    booktitle={2021 IEEE International Conference on Quantum Computing and Engineering (QCE)},
    publisher={IEEE},
    author={Burgholzer, Lukas and Bauer, Hartwig and Wille, Robert},
    year={2021},
    month=oct
}

@misc{goh2023liealgebraicclassicalsimulationsvariational,
      title={Lie-algebraic classical simulations for variational quantum computing}, 
      author={Matthew L. Goh and Martin Larocca and Lukasz Cincio and M. Cerezo and Frédéric Sauvage},
      year={2023},
      eprint={2308.01432},
      archivePrefix={arXiv},
      primaryClass={quant-ph},
      url={https://arxiv.org/abs/2308.01432}, 
}

@misc{tanaka2024quantumstatepreparationfree,
      title={Quantum State Preparation via Free Binary Decision Diagram}, 
      author={Yu Tanaka and Hayata Yamasaki and Mio Murao},
      year={2024},
      eprint={2407.01671},
      archivePrefix={arXiv},
      primaryClass={quant-ph},
      url={https://arxiv.org/abs/2407.01671}, 
}

@misc{kazi2024analyzingquantumapproximateoptimization,
      title={Analyzing the quantum approximate optimization algorithm: ans\"atze, symmetries, and Lie algebras}, 
      author={Sujay Kazi and Martín Larocca and Marco Farinati and Patrick J. Coles and M. Cerezo and Robert Zeier},
      year={2024},
      eprint={2410.05187},
      archivePrefix={arXiv},
      primaryClass={quant-ph},
      url={https://arxiv.org/abs/2410.05187}, 
}

@article{ragone2024lie,
	author = {Ragone, Michael and Bakalov, Bojko N. and Sauvage, Fr{\'e}d{\'e}ric and Kemper, Alexander F. and Ortiz Marrero, Carlos and Larocca, Mart{\'\i}n and Cerezo, M.},
	date = {2024/08/22},
	doi = {10.1038/s41467-024-49909-3},
	id = {Ragone2024},
	isbn = {2041-1723},
	journal = {Nature Communications},
	number = {1},
	pages = {7172},
	title = {A Lie algebraic theory of barren plateaus for deep parameterized quantum circuits},
	url = {https://doi.org/10.1038/s41467-024-49909-3},
	volume = {15},
	year = {2024}}

@inproceedings{burgholzer2022exploitingarbitrarypaths,
    author={Burgholzer, Lukas and Ploier, Alexander and Wille, Robert},
    booktitle={2022 Design, Automation \& Test in Europe Conference \& Exhibition}, 
    title={Exploiting Arbitrary Paths for the Simulation of Quantum Circuits with Decision Diagrams}, 
    year={2022},
    volume={},
    number={},
    pages={64-67},
    doi={10.23919/DATE54114.2022.9774631}
}

@article{grurl2023noiseaware,
    author={Grurl, Thomas and Fuß, Jürgen and Wille, Robert},
    journal={IEEE Transactions on Computer-Aided Design of Integrated Circuits and Systems}, 
    title={Noise-Aware Quantum Circuit Simulation With Decision Diagrams}, 
    year={2023},
    volume={42},
    number={3},
    pages={860-873},
    doi={10.1109/TCAD.2022.3182628}
}

@article{mcclean2018barren,
    author = {McClean, Jarrod R. and Boixo, Sergio and Smelyanskiy, Vadim N. and Babbush, Ryan and Neven, Hartmut},
    date = {2018/11/16},
    doi = {10.1038/s41467-018-07090-4},
    id = {McClean2018},
    isbn = {2041-1723},
    journal = {Nature Communications},
    number = {1},
    pages = {4812},
    title = {Barren plateaus in quantum neural network training landscapes},
    url = {https://doi.org/10.1038/s41467-018-07090-4},
    volume = {9},
    year = {2018},
}

@article{alcazar2024enhancing,
    author = {Alcazar, Javier and Ghazi Vakili, Mohammad and Kalayci, Can B. and Perdomo-Ortiz, Alejandro},
    date = {2024/03/29},
    doi = {10.1038/s41467-024-46959-5},
    id = {Alcazar2024},
    isbn = {2041-1723},
    journal = {Nature Communications},
    number = {1},
    pages = {2761},
    title = {Enhancing combinatorial optimization with classical and quantum generative models},
    url = {https://doi.org/10.1038/s41467-024-46959-5},
    volume = {15},
    year = {2024}
}

@article{verstraete2008matrix,
  title={Matrix product states, projected entangled pair states, and variational renormalization group methods for quantum spin systems},
  author={Verstraete, Frank and Murg, Valentin and Cirac, J Ignacio},
  journal={Advances in physics},
  volume={57},
  number={2},
  pages={143--224},
  year={2008},
  publisher={Taylor \& Francis}
}

@article{sharir2020deepautoregressive,
  title = {Deep Autoregressive Models for the Efficient Variational Simulation of Many-Body Quantum Systems},
  author = {Sharir, Or and Levine, Yoav and Wies, Noam and Carleo, Giuseppe and Shashua, Amnon},
  journal = {Phys. Rev. Lett.},
  volume = {124},
  issue = {2},
  pages = {020503},
  numpages = {6},
  year = {2020},
  month = {Jan},
  publisher = {American Physical Society},
  doi = {10.1103/PhysRevLett.124.020503},
  url = {https://link.aps.org/doi/10.1103/PhysRevLett.124.020503}
}

@inproceedings{wang1989training,
  title={On training of artificial neural networks},
  author={Wang and Malakooti},
  booktitle={International 1989 Joint Conference on Neural Networks},
  pages={387--393},
  year={1989},
  organization={IEEE}
}

@article{mazzola2024quantum,
  title={Quantum computing for chemistry and physics applications from a Monte Carlo perspective},
  author={Mazzola, Guglielmo},
  journal={The Journal of Chemical Physics},
  volume={160},
  number={1},
  year={2024},
  publisher={AIP Publishing}
}

@article{Bassman_Oftelie_2021,
doi = {10.1088/2058-9565/ac1ca6},
url = {https://dx.doi.org/10.1088/2058-9565/ac1ca6},
year = {2021},
month = {sep},
publisher = {IOP Publishing},
volume = {6},
number = {4},
pages = {043002},
author = {Bassman Oftelie, Lindsay and Urbanek, Miroslav and Metcalf, Mekena and Carter, Jonathan and Kemper, Alexander F and de Jong, Wibe A},
title = {Simulating quantum materials with digital quantum computers},
journal = {Quantum Science and Technology}
}

@article{meglio2024quantum,
  title = {Quantum Computing for High-Energy Physics: State of the Art and Challenges},
  author = {Di Meglio, Alberto and Jansen, Karl and Tavernelli, Ivano and Alexandrou, Constantia and Arunachalam, Srinivasan and Bauer, Christian W. and Borras, Kerstin and Carrazza, Stefano and Crippa, Arianna and Croft, Vincent and de Putter, Roland and Delgado, Andrea and Dunjko, Vedran and Egger, Daniel J. and Fern\'andez-Combarro, Elias and Fuchs, Elina and Funcke, Lena and Gonz\'alez-Cuadra, Daniel and Grossi, Michele and Halimeh, Jad C. and Holmes, Zo\"e and K\"uhn, Stefan and Lacroix, Denis and Lewis, Randy and Lucchesi, Donatella and Martinez, Miriam Lucio and Meloni, Federico and Mezzacapo, Antonio and Montangero, Simone and Nagano, Lento and Pascuzzi, Vincent R. and Radescu, Voica and Ortega, Enrique Rico and Roggero, Alessandro and Schuhmacher, Julian and Seixas, Joao and Silvi, Pietro and Spentzouris, Panagiotis and Tacchino, Francesco and Temme, Kristan and Terashi, Koji and Tura, Jordi and T\"uys\"uz, Cenk and Vallecorsa, Sofia and Wiese, Uwe-Jens and Yoo, Shinjae and Zhang, Jinglei},
  journal = {PRX Quantum},
  volume = {5},
  issue = {3},
  pages = {037001},
  numpages = {49},
  year = {2024},
  month = {Aug},
  publisher = {American Physical Society},
  doi = {10.1103/PRXQuantum.5.037001},
  url = {https://link.aps.org/doi/10.1103/PRXQuantum.5.037001}
}

@article{tindall2024efficient,
  title = {Efficient Tensor Network Simulation of IBM's Eagle Kicked Ising Experiment},
  author = {Tindall, Joseph and Fishman, Matthew and Stoudenmire, E. Miles and Sels, Dries},
  journal = {PRX Quantum},
  volume = {5},
  issue = {1},
  pages = {010308},
  numpages = {16},
  year = {2024},
  month = {Jan},
  publisher = {American Physical Society},
  doi = {10.1103/PRXQuantum.5.010308},
  url = {https://link.aps.org/doi/10.1103/PRXQuantum.5.010308}
}

@article{patra2024efficient,
  title = {Efficient tensor network simulation of IBM's largest quantum processors},
  author = {Patra, Siddhartha and Jahromi, Saeed S. and Singh, Sukhbinder and Or\'us, Rom\'an},
  journal = {Phys. Rev. Res.},
  volume = {6},
  issue = {1},
  pages = {013326},
  numpages = {7},
  year = {2024},
  month = {Mar},
  publisher = {American Physical Society},
  doi = {10.1103/PhysRevResearch.6.013326},
  url = {https://link.aps.org/doi/10.1103/PhysRevResearch.6.013326}
}

@article{carleo2017solving,
  title={Solving the quantum many-body problem with artificial neural networks},
  author={Carleo, Giuseppe and Troyer, Matthias},
  journal={Science},
  volume={355},
  number={6325},
  pages={602--606},
  year={2017},
  publisher={American Association for the Advancement of Science}
}

@Article{netket3,
    title={{NetKet 3: Machine Learning Toolbox for Many-Body Quantum Systems}},
    author={Filippo Vicentini and Damian Hofmann and Attila Szabó and Dian Wu and Christopher   Roth and Clemens Giuliani and Gabriel Pescia and Jannes Nys and Vladimir Vargas-Calderón and   Nikita Astrakhantsev and Giuseppe Carleo},
    journal={SciPost Phys. Codebases},
    pages={7},
    year={2022},
    publisher={SciPost},
    doi={10.21468/SciPostPhysCodeb.7},
    url={https://scipost.org/10.21468/SciPostPhysCodeb.7},
}

@misc{paszke2019pytorchimperativestylehighperformance,
    title={PyTorch: An Imperative Style, High-Performance Deep Learning Library}, 
    author={Adam Paszke and Sam Gross and Francisco Massa and Adam Lerer and James Bradbury and Gregory Chanan and Trevor Killeen and Zeming Lin and Natalia Gimelshein and Luca Antiga and Alban Desmaison and Andreas Köpf and Edward Yang and Zach DeVito and Martin Raison and Alykhan Tejani and Sasank Chilamkurthy and Benoit Steiner and Lu Fang and Junjie Bai and Soumith Chintala},
    year={2019},
    eprint={1912.01703},
    archivePrefix={arXiv},
    primaryClass={cs.LG},
    url={https://arxiv.org/abs/1912.01703}, 
}

@software{jax2018github,
    author = {James Bradbury and Roy Frostig and Peter Hawkins and Matthew James Johnson and Chris Leary and Dougal Maclaurin and George Necula and Adam Paszke and Jake Vander{P}las and Skye Wanderman-{M}ilne and Qiao Zhang},
    title = {{JAX}: composable transformations of {P}ython+{N}um{P}y programs},
    url = {http://github.com/google/jax},
    version = {0.3.13},
    year = {2018},
}

@article{anschuetz2022traps,
    author = {Anschuetz, Eric R. and Kiani, Bobak T.},
    date = {2022/12/15},
    doi = {10.1038/s41467-022-35364-5},
    id = {Anschuetz2022},
    isbn = {2041-1723},
    journal = {Nature Communications},
    number = {1},
    pages = {7760},
    title = {Quantum variational algorithms are swamped with traps},
    url = {https://doi.org/10.1038/s41467-022-35364-5},
    volume = {13},
    year = {2022}
}

@misc{bermejo2024quantumconvolutionalneuralnetworks,
    title={Quantum Convolutional Neural Networks are (Effectively) Classically Simulable}, 
    author={Pablo Bermejo and Paolo Braccia and Manuel S. Rudolph and Zoë Holmes and Lukasz Cincio and M. Cerezo},
    year={2024},
    eprint={2408.12739},
    archivePrefix={arXiv},
    primaryClass={quant-ph},
    url={https://arxiv.org/abs/2408.12739}, 
}

@misc{gilfuster2024relationtrainabilitydequantizationvariational,
    title={On the relation between trainability and dequantization of variational quantum learning models}, 
    author={Elies Gil-Fuster and Casper Gyurik and Adrián Pérez-Salinas and Vedran Dunjko},
    year={2024},
    eprint={2406.07072},
    archivePrefix={arXiv},
    primaryClass={quant-ph},
    url={https://arxiv.org/abs/2406.07072}, 
}

@article{sun2020optimization,
    author = {Sun, Ruo-Yu},
    date = {2020/06/01},
    doi = {10.1007/s40305-020-00309-6},
    id = {Sun2020},
    isbn = {2194-6698},
    journal = {Journal of the Operations Research Society of China},
    number = {2},
    pages = {249--294},
    title = {Optimization for Deep Learning: An Overview},
    url = {https://doi.org/10.1007/s40305-020-00309-6},
    volume = {8},
    year = {2020}
}

@article{vargas2021tsp,
    author = {Vargas-Calder\'{o}n ,Vladimir and Parra-A. ,Nicolas and Vinck-Posada ,Herbert and Gonz\'{a}lez ,Fabio A.},
    doi = {10.7566/JPSJ.90.114002},
    eprint = {https://doi.org/10.7566/JPSJ.90.114002},
    journal = {Journal of the Physical Society of Japan},
    number = {11},
    pages = {114002},
    title = {Many-Qudit Representation for the Travelling Salesman Problem Optimisation},
    url = {https://doi.org/10.7566/JPSJ.90.114002},
    volume = {90},
    year = {2021}
}

@misc{kingma2017adam,
      title={Adam: A Method for Stochastic Optimization}, 
      author={Diederik P. Kingma and Jimmy Ba},
      year={2017},
      eprint={1412.6980},
      archivePrefix={arXiv},
      primaryClass={cs.LG},
      url={https://arxiv.org/abs/1412.6980}, 
}

@misc{vivas2022nqs,
    title={Neural-Network Quantum States: A Systematic Review}, 
    author={David R. Vivas and Javier Madroñero and Victor Bucheli and Luis O. Gómez and John H. Reina},
    year={2022},
    eprint={2204.12966},
    archivePrefix={arXiv},
    primaryClass={quant-ph},
    url={https://arxiv.org/abs/2204.12966}, 
}

@article{schollwock2011mps,
   title={The density-matrix renormalization group in the age of matrix product states},
   volume={326},
   ISSN={0003-4916},
   url={http://dx.doi.org/10.1016/j.aop.2010.09.012},
   DOI={10.1016/j.aop.2010.09.012},
   number={1},
   journal={Annals of Physics},
   publisher={Elsevier BV},
   author={Schollwöck, Ulrich},
   year={2011},
   month=jan, pages={96–192}
}

@article{orus2014tensornetworks,
   title={A practical introduction to tensor networks: Matrix product states and projected entangled pair states},
   volume={349},
   ISSN={0003-4916},
   url={http://dx.doi.org/10.1016/j.aop.2014.06.013},
   DOI={10.1016/j.aop.2014.06.013},
   journal={Annals of Physics},
   publisher={Elsevier BV},
   author={Orús, Román},
   year={2014},
   month=oct, pages={117–158}
}

@misc{rudolph2025paulipropagationcomputationalframework,
      title={Pauli Propagation: A Computational Framework for Simulating Quantum Systems}, 
      author={Manuel S. Rudolph and Tyson Jones and Yanting Teng and Armando Angrisani and Zoë Holmes},
      year={2025},
      eprint={2505.21606},
      archivePrefix={arXiv},
      primaryClass={quant-ph},
      url={https://arxiv.org/abs/2505.21606}, 
}

@article{Fargier_Marquis_Niveau_Schmidt_2014,
	author = {Fargier, H{\'e}l{\`e}ne and Marquis, Pierre and Niveau, Alexandre and Schmidt, Nicolas},
	doi = {10.1609/aaai.v28i1.8853},
	journal = {Proceedings of the AAAI Conference on Artificial Intelligence},
	month = {Jun.},
	number = {1},
	title = {A Knowledge Compilation Map for Ordered Real-Valued Decision Diagrams},
	url = {https://ojs.aaai.org/index.php/AAAI/article/view/8853},
	volume = {28},
	year = {2014}}

@article{sistla2024cflobdds,
author = {Sistla, Meghana Aparna and Chaudhuri, Swarat and Reps, Thomas},
title = {CFLOBDDs: Context-Free-Language Ordered Binary Decision Diagrams},
year = {2024},
issue_date = {June 2024},
publisher = {Association for Computing Machinery},
address = {New York, NY, USA},
volume = {46},
number = {2},
issn = {0164-0925},
url = {https://doi.org/10.1145/3651157},
doi = {10.1145/3651157},
journal = {ACM Trans. Program. Lang. Syst.},
month = may,
articleno = {7},
numpages = {82}
}

@inproceedings{Kisa2014PSDD,
  author    = {Doga Kisa and Guy Van den Broeck and Arthur Choi and Adnan Darwiche},
  title     = {Probabilistic Sentential Decision Diagrams},
  booktitle = {Proceedings of the Fourteenth International Conference on Principles of Knowledge Representation and Reasoning (KR)},
  year      = {2014},
  pages     = {558--567},
  publisher = {AAAI Press},
  address   = {Vienna, Austria}
}

@inproceedings{Wilson2005Semiring,
  author    = {Nic Wilson},
  title     = {Decision Diagrams for the Computation of Semiring Valuations},
  booktitle = {Proceedings of the 19th International Joint Conference on Artificial Intelligence (IJCAI)},
  year      = {2005},
  pages     = {331--336},
  address   = {Edinburgh, Scotland},
  publisher = {Morgan Kaufmann Publishers}
}

@inproceedings{thierry2009hierarchical,
	address = {Berlin, Heidelberg},
	author = {Thierry-Mieg, Yann and Poitrenaud, Denis and Hamez, Alexandre and Kordon, Fabrice},
	booktitle = {Tools and Algorithms for the Construction and Analysis of Systems},
	editor = {Kowalewski, Stefan and Philippou, Anna},
	isbn = {978-3-642-00768-2},
	pages = {1--15},
	publisher = {Springer Berlin Heidelberg},
	title = {Hierarchical Set Decision Diagrams and Regular Models},
	year = {2009}}

@inproceedings{Sanner2005AADD,
  author    = {Scott Sanner and David McAllester},
  title     = {Affine Algebraic Decision Diagrams (AADDs) and Their Application to Structured Probabilistic Inference},
  booktitle = {Proceedings of the 19th International Joint Conference on Artificial Intelligence (IJCAI)},
  year      = {2005},
  pages     = {1384--1390},
  address   = {Edinburgh, Scotland},
  publisher = {Morgan Kaufmann Publishers}
}

@article{Fujita1997MTBDD,
  author    = {M. Fujita and P. C. McGeer and J. C.-Y. Yang},
  title     = {Multi‐Terminal Binary Decision Diagrams: An Efficient Data Structure for Matrix Representation},
  journal   = {Formal Methods in System Design},
  volume    = {10},
  number    = {2-3},
  pages     = {149--169},
  year      = {1997},
  doi       = {10.1023/A:1008647823331}
}

@article{Lai1996EVBDD,
  author    = {Yung‐Te Lai and Massoud Pedram and Sarma B. K. Vrudhula},
  title     = {Formal Verification using Edge-Valued Binary Decision Diagrams},
  journal   = {IEEE Transactions on Computers},
  volume    = {45},
  number    = {2},
  pages     = {247--255},
  year      = {1996},
  doi       = {10.1109/12.485378}
}

@article{Tafertshofer1997FEVBDD,
  author    = {P. Tafertshofer and Massoud Pedram},
  title     = {Factored Edge-Valued Binary Decision Diagrams},
  journal   = {Formal Methods in System Design},
  volume    = {10},
  number    = {2},
  pages     = {243--270},
  year      = {1997},
  doi       = {10.1023/A:1008691605584}
}

@INPROCEEDINGS{miller2006qmdd,
  author={Miller, D.M. and Thornton, M.A.},
  booktitle={36th International Symposium on Multiple-Valued Logic (ISMVL'06)}, 
  title={QMDD: A Decision Diagram Structure for Reversible and Quantum Circuits}, 
  year={2006},
  volume={},
  number={},
  pages={30-30},
  doi={10.1109/ISMVL.2006.35}}

@misc{vinkhuijzen2024knowledgecompilationmapquantum,
      title={A Knowledge Compilation Map for Quantum Information}, 
      author={Lieuwe Vinkhuijzen and Tim Coopmans and Alfons Laarman},
      year={2024},
      eprint={2401.01322},
      archivePrefix={arXiv},
      primaryClass={quant-ph},
      url={https://arxiv.org/abs/2401.01322}, 
}

@article{sistla2024weighted,
author = {Sistla, Meghana and Chaudhuri, Swarat and Reps, Thomas},
title = {Weighted Context-Free-Language Ordered Binary Decision Diagrams},
year = {2024},
issue_date = {October 2024},
publisher = {Association for Computing Machinery},
address = {New York, NY, USA},
volume = {8},
number = {OOPSLA2},
url = {https://doi.org/10.1145/3689760},
doi = {10.1145/3689760},
journal = {Proc. ACM Program. Lang.},
month = oct,
articleno = {320},
numpages = {30}
}

@article{Burgholzer2025MQTCore,
  author    = {Lukas Burgholzer and Yannick Stade and Tom Peham and Robert Wille},
  title     = {MQT Core: The Backbone of the Munich Quantum Toolkit (MQT)},
  journal   = {Journal of Open Source Software},
  year      = {2025},
  volume    = {10},
  number    = {108},
  pages     = {7478},
  publisher = {The Open Journal},
  doi       = {10.21105/joss.07478},
  url       = {https://doi.org/10.21105/joss.07478}
}

@misc{hong2025limtddcompactdecisiondiagram,
      title={LimTDD: A Compact Decision Diagram Integrating Tensor and Local Invertible Map Representations}, 
      author={Xin Hong and Aochu Dai and Dingchao Gao and Sanjiang Li and Zhengfeng Ji and Mingsheng Ying},
      year={2025},
      eprint={2504.01168},
      archivePrefix={arXiv},
      primaryClass={cs.DS},
      url={https://arxiv.org/abs/2504.01168}, 
}

@misc{brand2025qsylvanparalleldecisiondiagram,
      title={Q-Sylvan: A Parallel Decision Diagram Package for Quantum Computing}, 
      author={Sebastiaan Brand and Alfons Laarman},
      year={2025},
      eprint={2508.00514},
      archivePrefix={arXiv},
      primaryClass={quant-ph},
      url={https://arxiv.org/abs/2508.00514}, 
}

@article{cerezo2021variational,
	author = {Cerezo, M. and Arrasmith, Andrew and Babbush, Ryan and Benjamin, Simon C. and Endo, Suguru and Fujii, Keisuke and McClean, Jarrod R. and Mitarai, Kosuke and Yuan, Xiao and Cincio, Lukasz and Coles, Patrick J.},
	date = {2021/09/01},
	doi = {10.1038/s42254-021-00348-9},
	id = {Cerezo2021},
	isbn = {2522-5820},
	journal = {Nature Reviews Physics},
	number = {9},
	pages = {625--644},
	title = {Variational quantum algorithms},
	url = {https://doi.org/10.1038/s42254-021-00348-9},
	volume = {3},
	year = {2021}}

@article{nakata2021quantum,
  title = {Quantum Circuits for Exact Unitary $t$-Designs and Applications to Higher-Order Randomized Benchmarking},
  author = {Nakata, Yoshifumi and Zhao, Da and Okuda, Takayuki and Bannai, Eiichi and Suzuki, Yasunari and Tamiya, Shiro and Heya, Kentaro and Yan, Zhiguang and Zuo, Kun and Tamate, Shuhei and Tabuchi, Yutaka and Nakamura, Yasunobu},
  journal = {PRX Quantum},
  volume = {2},
  issue = {3},
  pages = {030339},
  numpages = {35},
  year = {2021},
  month = {Sep},
  publisher = {American Physical Society},
  doi = {10.1103/PRXQuantum.2.030339},
  url = {https://link.aps.org/doi/10.1103/PRXQuantum.2.030339}
}

@article{belkin2024approximate,
  title = {Approximate $t$-Designs in Generic Circuit Architectures},
  author = {Belkin, Daniel and Allen, James and Ghosh, Soumik and Kang, Christopher and Lin, Sophia and Sud, James and Chong, Frederic T. and Fefferman, Bill and Clark, Bryan K.},
  journal = {PRX Quantum},
  volume = {5},
  issue = {4},
  pages = {040344},
  numpages = {26},
  year = {2024},
  month = {Dec},
  publisher = {American Physical Society},
  doi = {10.1103/PRXQuantum.5.040344},
  url = {https://link.aps.org/doi/10.1103/PRXQuantum.5.040344}
}

@article{ran2020encoding,
  title = {Encoding of matrix product states into quantum circuits of one- and two-qubit gates},
  author = {Ran, Shi-Ju},
  journal = {Phys. Rev. A},
  volume = {101},
  issue = {3},
  pages = {032310},
  numpages = {7},
  year = {2020},
  month = {Mar},
  publisher = {American Physical Society},
  doi = {10.1103/PhysRevA.101.032310},
  url = {https://link.aps.org/doi/10.1103/PhysRevA.101.032310}
}

@article{rudolph2023decomposition,
  title={Decomposition of matrix product states into shallow quantum circuits},
  author={Rudolph, Manuel S and Chen, Jing and Miller, Jacob and Acharya, Atithi and Perdomo-Ortiz, Alejandro},
  journal={Quantum Science and Technology},
  volume={9},
  number={1},
  pages={015012},
  year={2023},
  publisher={IOP Publishing}
}

@ARTICLE{qi2024variational,
       author = {{Qi}, Han and {Xiao}, Sihui and {Liu}, Zhuo and {Gong}, Changqing and {Gani}, Abdullah},
        title = "{Variational quantum algorithms: fundamental concepts, applications and challenges}",
      journal = {Quantum Information Processing},
         year = 2024,
        month = jun,
       volume = {23},
       number = {6},
          eid = {224},
        pages = {224},
          doi = {10.1007/s11128-024-04438-2},
       adsurl = {https://ui.adsabs.harvard.edu/abs/2024QuIP...23..224Q}
}
\clearpage
\appendix
\section{Variational Monte Carlo}\label{sec:vmc}

Variational Monte Carlo (VMC) is a probabilistic algorithm for estimating the ground state of a Hamiltonian based on the variational principle, where
\begin{align}
     \expval{H}{\psi_{\vb*{\theta}}} \ge E_0,
\end{align}
where $E_0$ is the true ground state energy of the Hamiltonian $H$.
This appendix is a condensed review of VMC applied to Hamiltonian ground state estimation in the context of ansätze defined by architectures that can be autodifferentiated.
We suggest that the interested reader reviews Ref. \cite{netket3}, which contains a more in-depth explanation of VMC in this domain.
The algorithm consists on iteratively updating the parameters $\vb*{\theta}$ of any ansatz---in our case, the VDD---by going through the following steps:
\begin{itemize}
    \item Build a sample $\{\vb*{b}\}$, where $\vb*{b}\sim\abs{\psi_{\vb*{\theta}}}^2$, i.e., a sample of bit strings that follow the Born distribution of the variational state defined by the VDD.
    \item Estimate $\grad_{\vb*{\theta}} \expval{H}{\psi_{\vb*{\theta}}}$ with respect to the previously built sample.
    \item Use gradient-descent rules to update the variational parameters $\vb*{\theta}$, reaching a lower value of $\expval{H}{\psi_{\vb*{\theta}}}$ with respect to the built sample.
\end{itemize}
These steps can be repeated as many times as it is necessary to converge the expected value of the Hamiltonian.
In what follows, we explain each step.
\subsection{Building bit string samples}\label{sec:autoregressive_sampling}

Unlike many neural quantum state architectures that require the Metropolis-Hastings algorithm for sampling \cite{netket3}, exact sampling can be performed directly in VDD, resulting in an unbiased sample \cite{vivas2022nqs}.
The process of exact sampling is straightforward.
Starting at the root node, the value of each qubit is sampled sequentially.
The first qubit is sampled as $\ket{0}$ with a probability of $r^2$, where $r$ is the amplitude parameter of the corresponding node, or as $\ket{1}$ with a probability of $1 - r^2$.
After determining the state of the qubit, the corresponding edge is selected, leading to the next node that represents the following qubit.
The state of this qubit is then sampled using the respective $r$ parameter.
This process continues until the states of all qubits are sampled, ultimately yielding an unbiased sample from the VDD.

This process allows us to perform exact sampling from the Born distribution induced by the VDD, i.e., we can sample bit strings $\vb*{b}\sim p(\vb*{b})=\abs{\psi_{\vb*{\theta}}(\vb*{b})}^2$.

\added{Similar to other classical ansätze such as matrix product states (MPS), one can deploy them on quantum circuits for faster sampling.
Sampling autoregressively from MPSs or VDDs has a time complexity of $\mathcal{O}(n)$, whereas sampling from the same quantum state using a quantum computer depends on the circuit depth of the quantum circuit that realises such ansatz. 
This depth can depend on the number of qubits $n$ in different ways, depending on the algorithm used for synthesising a quantum circuit from one of these ansätze.
For MPSs, some examples of conversion algorithms are Refs.~\citep{ran2020encoding,rudolph2023decomposition}.
Therefore, mapping the VDD ansatz to an explicit quantum circuit could be of interest as a means of accelerating bit string sampling.
This is left for future work.}

\subsection{Estimating expected values of observables and their derivatives}

The expected value of any observable $A$ with respect to the VDD can be computed as~\cite{netket3}
\begin{align}
    \expval{A} = \mathbb{E}[\tilde{A}] = \sum_{\vb*{b}}p(\vb*{b})\tilde{A}(\vb*{b}),
\end{align}
where $\tilde{A}$ is the local estimator of $A$, defined as
\begin{align}
    \tilde{A}(\vb*{b}) = \sum_{\vb*{b}'}\frac{\psi_{\vb*{\theta}}(\vb*{b}')}{\psi_{\vb*{\theta}}(\vb*{b})}\bra{\vb*{b}}A\ket*{\vb*{b}'},\label{eq:localestimator}
\end{align}
where the sum runs over all elements of the Hilbert space basis.
Note that the sum in \eqref{eq:localestimator} contains at most $k$ terms for a $k$-local observable $A$.
This means that even when the sum is over all the elements of the Hilbert space, the sum is tractable and efficiently carried out due to the sparsity of $A$.

The derivative of the expected value of $A$ with respect to a parameter\footnote{As in \cref{eq:variance_of_gradients}, we use $j$ here to refer to a single parameter of the VDD, not to the collection of three parameters of the $j$-th node of the VDD shown in~\cref{fig:accordion}.} of the VDD can be written as~\cite{netket3}
\begin{align}
    \pdv{\expval{A}}{\theta_j} = 2\Re(\mathbb{E}[O_j^*(\tilde{A} - \mathbb{E}[\tilde{A}])]),
\end{align}
where $O_j$ is given by
\begin{align}
    O_j=\pdv{\log\psi_{\vb*{\theta}}(\vb*{b})}{\theta_j}.\label{eq:logderivativepsi}
\end{align}
$O_j$ is particularly simple to calculate in the VDD.
As noted in the main text (cf.\cref{eq:element_of_vdd}), $\psi_{\vb*{\theta}}(\vb*{b})$ is a product of complex numbers written as $\alpha_i^{\beta_i}$, where $\alpha_i$ is a function of the amplitude parameter $r_i$ of a node $i$ in the VDD, and $\beta_i$ is a function of the phase parameters $\omega_i$ and $\phi_i$ of this node.
The exact dependence of $\alpha_i$ and $\beta_i$ on each node's parameters $(r_i,\omega_i,\phi_i)$ is determined by the nodes that are traversed as specified by the bit string $\vb*{b}$.
The logarithm in \cref{eq:logderivativepsi} further simplifies the differentiation, because one needs to take the derivative of a term that looks like
\begin{align}
    \sum_{i=1}^n \log \alpha_i^{\beta_i}
\end{align}
with respect to a parameter that appears only in one of the summands.

The final ingredient for estimating the expected value of any observable $A$ and its derivative with respect to the variational parameters is noticing that the expected values can be taken with respect to a sample of bit strings sampled from the Born distribution induced by the VDD.

\subsection{Updating variational parameters}

To close the circle, we are now able to update the parameters in a convenient way so as to minimise the expected value of the Hamiltonian $H$.
This can be achieved by several gradient descent algorithms \cite{sun2020optimization}, the simpler of which can be given by the stochastic gradient descent update rule:
\begin{align}
    \theta_j \leftarrow \theta_j - \eta \pdv{\expval{H}}{\theta_j},
\end{align}
where the derivative is estimated through Monte Carlo, and $\eta$ is referred to as the learning rate.

We highlight the fact that variational Monte Carlo setups allow tackling problems of hundreds of qubits with very limited computational resources, such as a personal laptop~\cite{vargas2021tsp}.

\end{document}